\documentclass[english,russian]{epl}
\usepackage{amsfonts}
\usepackage{amsthm}
\usepackage[hypertex]{hyperref}
\newtheorem{thm}{Theorem}

\newtheorem{lem}[thm]{Lemma}

\theoremstyle{definition}
\newtheorem{defn}[thm]{Definition}
\theoremstyle{remark}
\newtheorem{example}[thm]{Example}

\providecommand{\arXiv}[1]{\eprint{http://arXiv.org/abs/#1}{#1}}
\providecommand{\algebra}[1]{\ensuremath{\mathfrak{#1}}}
\providecommand{\object}[2][\,]{\ensuremath{\mathrm{#2}#1}}
\providecommand{\Space}[3][]{\ensuremath{\mathbb{#2}^{#3}_{#1}{}}}
\providecommand{\such}{\,\mid\,}
\providecommand{\FSpace}[3][]{\ensuremath{#2_{#3}^{#1}{}}} 
\providecommand{\norm}[2][\relax]{\left\|#2\right\|\ifx#1\relax\else_{#1}\fi}
\providecommand{\modulus}[2][\relax]{\left| #2 \right|\ifx#1\relax\else_{#1}\fi}
\providecommand{\intersect}[2]{\left.#1\right|_{#2}}
\providecommand{\eqref}[1]{\textup{(\ref{#1})}}
\providecommand{\comment}[1]{}
\providecommand{\href}[2]{#2}
  \providecommand{\MR}[1]{\textbf{MR}~\href{http://www.ams.org/mathscinet-getitem?mr=#1}{\#~#1}}
\providecommand{\eprint}[2]{\href{#1}{\texttt{#2}}}
\def\ifundefined#1{\expandafter\ifx\csname#1\endcsname\relax}

\providecommand{\Zbl}[1]{\textbf{Zbl}~\href{http://www.emis.de:80/cgi-bin/zmen/ZMATH/en/zmathf.html?first=1&maxdocs=3&type=html&an=#1&format=complete}{\#~#1}}
\providecommand{\myhbar}{h}
\providecommand{\rmi}{\mathrm{i}}
\providecommand{\orbit}[1]{\mathcal{O}_{#1}}
\providecommand{\uir}[1]{\rho_{#1}}

\providecommand{\anti}{\mathcal{A}}

\providecommand{\ub}[3][]{\left\{\!#1\left[#2,#3\right]\!#1\right\}}

\providecommand{\fl}{}

\providecommand{\keywords}[1]{}
\providecommand{\urladdr}[1]{}
\RequirePackage[T2A]{fontenc}
\usepackage[koi8-r]{inputenc}   
\usepackage[british,russian]{babel}

\title{A Quantum--Classical Bracket from $p$-Mechanics}
\author
{\href{http://maths.leeds.ac.uk/~kisilv/}{Vladimir
    V. Kisil}}

\institute{
School of Mathematics,
University of Leeds,
Leeds LS2\,9JT,
UK \\
On leave from Odessa University}
\begin{document}


\maketitle

\begin{abstract}
  We provide an answer to the long standing problem of mixing quantum
  and classical dynamics within a single formalism. The construction
  is based on \(p\)-mechanical derivation of quantum and classical
  dynamics from the representation theory of the Heisenberg group. To
  achieve a quantum-classical mixing we take the product of two copies
  of the Heisenberg group which represent two
  different Planck's constants. In comparison with earlier guesses our
  answer contains an extra term of analytical nature, which was not
  obtained before in purely algebraic setup.
\end{abstract}
\hfill{\small В одну телегу впрячь неможно коня и трепетную лань.
  \emph{А.С. Пушкин}}

\section{Introduction}
\label{sec:introduction}

There is a strong and persistent interest over decades in \emph{a
  self-consistent model for an aggregate system, which combines
  components with both quantum and classical behaviour}
(see~\cite{Aleksandrov81,BoucherTaschen88,Anderson97a,Prezhdo-Kisil97,CaroSalcedo99,DiosiGisinStrunz00a,Sahoo04}
and references therein). There are various reasons for such an
interest.  Firstly, there are many questions where considerations of
quantum-classical aggregates are unavoidable, e.g. measurement of a quantum
system by a classical apparatus, a quantum particle in the classical
gravity field, etc. Secondly, even for purely quantum conglomerates we
expect that a quantum-classical approximation may be easier for
investigation than the purely quantum picture. Thus it is natural that
models of quantum-classical interaction became of separate theoretical
interest. 

The discussion is typically linked to a search of \emph{quantum-classical
bracket} which should combine properties of the quantum commutator
\([\cdot, \cdot]\) and Poisson's bracket \(\{\cdot, \cdot\}\) in the
corresponding sectors. Some simple algebraic combinations like
\begin{equation}
  \label{eq:aleksandrov-brackets}
  \frac{1}{\rmi\myhbar}[A,B] +\frac{1}{2}(\{A,B\}-\{B,A\})
\end{equation}
were guessed during the last twenty five
years~\cite{Aleksandrov81,BoucherTaschen88,Anderson97a} but neither of
them turned to be completely satisfactory. Moreover several ``no-go''
theorems in that direction were proved over the last ten
years~\cite{Salcedo96,CaroSalcedo99,Sahoo04}. Thus the prevailing
opinion now is that no consistent quantum-classical
bracket is possible. However the explicit similarity between the
Hamiltonian descriptions of quantum and classical dynamics repeatedly
undermine such a believe.

This paper builds a consistent quantum-classical bracket within the
framework of
\(p\)-mechanics~\cite{Kisil96a,Kisil00a,Kisil02e,Kisil04a}.  This
approach is based on the representation theory of nilpotent Lie groups
(the Heisenberg group \(\Space{H}{n}\) in the first instance) and
naturally embeds both quantum and classical descriptions.
\(p\)-Mechanical observables are convolutions on a nilpotent group
\(G\) and contain both classical and quantum pictures for all values of
Planck's constants at the same time. These pictures can be separated
by a restriction of \(p\)-observables to irreducible representations
of \(G\), e.g. by considering their actions on \(p\)-mechanical
states~\cite{Brodlie03a,BrodlieKisil03a}.

The important step~\cite{Kisil00a,Kisil02e} is the definition of the
universal bracket between convolutions on the Heisenberg group, which
are transformed by the above mentioned representations into the
quantum commutator (Moyal bracket~\cite{Zachos02a}) and the Poisson
bracket correspondingly. Consequently it is sufficient to solve the
dynamic equation in \(p\)-mechanics in order to obtain both quantum
and classical dynamics. Since the universal bracket is based on the
usual commutator of convolutions (i.e. inner derivations of the
convolution algebra) they satisfy all important requirements, i.e.
linearity, antisymmetry, Leibniz and Jacoby
identities~\cite{CaroSalcedo99}. Moreover due to presence of
antiderivative operator~\eqref{eq:def-anti} the universal bracket
with a Hamiltonian has the dimensionality of time
derivative~\cite{Kisil02e}.  This approach was extended to quantum
field theory in~\cite{Kisil04a}. A brief account of
\(p\)-mechanics is provided in the first part of this paper.

To construct quantum-classical bracket we develop \(p\)-mechanics on
the group \(\Space{D}{n}\), which is the product of two copies of the
Heisenberg group \(\Space{H}{n}\). The group \(\Space{D}{n}\) was
already used to this end in our earlier paper~\cite{Prezhdo-Kisil97}
but the right bracket was not derived there due to several reasons:
the derivation followed the notorious semiclassical limit procedure;
the universal bracket~\cite{Kisil00a,Kisil02e} was not known at that
time. A correct derivation of quantum-classical bracket in the
consistent \(p\)-mechanical framework is given in the second part of
the present note. This bracket~\eqref{eq:qc-brackets} includes as a
part the Aleksandrov's bracket~\eqref{eq:aleksandrov-brackets}
together with an extra term of analytical nature, which involves
derivative with respect to the second Planck's constant,
see~\eqref{eq:qc-brackets}. This analytic term escapes all previous
purely algebraic considerations and ``no-go''
theorems~\cite{Salcedo96,CaroSalcedo99,Sahoo04} for the obvious
reasons.

Future investigations of these new quantum-classical bracket will
be given elsewhere.

\section{The Heisenberg group and $p$-mechanical bracket}
\label{sec:heisenberg-group-p}

\subsection{The Heisenberg group and its representations}
\label{sec:preliminaries}

Let \((s,x,y)\), where \(x\), \(y\in \Space{R}{n}\) and \(s\in\Space{R}{}\), be
an element of the Heisenberg group
\(\Space{H}{n}\)~\cite{Folland89,Howe80b}. The group law on
\(\Space{H}{n}\) is given as follows:
\begin{equation}
  \label{eq:H-n-group-law}
  \textstyle
  (s,x,y)*(s',x',y')=(s+s'+\frac{1}{2}\omega(x,y;x',y'),x+x',y+y'), 
\end{equation} 
where the non-commutativity is due to \(\omega\)---the
\emph{symplectic form} \(\omega(x,y;x',y')=xy'-x'y\) on
\(\Space{R}{2n}\)~\cite[\S~37]{Arnold91}.  The Lie algebra
\(\algebra{h}^n\) of \(\Space{H}{n}\) is spanned by left-invariant
vector fields
\begin{equation}
\textstyle  S={\partial_s}, \quad
  X_j=\partial_{ x_j}-\frac{1}{2}y_j{\partial_s},  \quad
 Y_j=\partial_{y_j}+\frac{1}{2}x_j{\partial_s}
  \label{eq:h-lie-algebra}
\end{equation}
on \(\Space{H}{n}\) with the Heisenberg \emph{commutator relations} 
 \( [X_i,Y_j]=\delta_{i,j}S\) 
and  all other commutators vanishing. 
There is the \emph{co-adjoint
  representation}~\cite[\S~15.1]{Kirillov76} \(\object{Ad}^*:
\algebra{h}^*_n \rightarrow \algebra{h}^*_n\) of \(\Space{H}{n}\):
\begin{equation}
  \label{eq:co-adjoint-rep}
  \object{ad}^*(s,x,y): (\myhbar ,q,p) \mapsto (\myhbar , q+\myhbar y,
  p-\myhbar x), \quad
\end{equation}
where \((\myhbar ,q,p)\in\algebra{h}^*_n\) in bi-orthonormal
coordinates to the exponential ones on \(\algebra{h}^n\).  There are
two types of orbits in~\eqref{eq:co-adjoint-rep} for
\(\object{Ad}^*\), i.e. Euclidean spaces \(\Space{R}{2n}\) and single
points:
\begin{eqnarray}
  \label{eq:co-adjoint-orbits-inf}
  \orbit{\myhbar} & = & \{(\myhbar, q,p): \textrm{ for 
  }\myhbar\neq 0,  (q,p) \in  \Space{R}{2n}\}, \qquad 
  \label{eq:co-adjoint-orbits-one}
  \orbit{(q,p)}  =  \{(0,q,p): \textrm{ for 
  } (q,p)\in \Space{R}{2n}\}.
\end{eqnarray} 
All representations are \emph{induced}~\cite[\S~13]{Kirillov76} by a
character \(\chi_\myhbar(s,0,0)=e^{2\pi \rmi \myhbar s}\) of the
centre of \(\Space{H}{n}\) generated by
\((\myhbar,0,0)\in\algebra{h}^*_n\) and
shifts~\eqref{eq:co-adjoint-rep} from the \emph{left} on
orbits~\eqref{eq:co-adjoint-orbits-inf}.  The explicit
formula respecting \emph{physical units}~\cite{Kisil02e} is:
\begin{equation}
  \textstyle
  \label{eq:stone-inf}
  \uir{\myhbar}(s,x,y): f_\myhbar (q,p) \mapsto 
  e^{ -2\pi\rmi( \myhbar s+qx+py)}
  f_\myhbar (q-\frac{\myhbar}{2} y, p+\frac{\myhbar}{2} x).
\end{equation}
The Stone--von Neumann theorem~\textup{\cite[\S~18.4]{Kirillov76}, \cite[Chap.~1,
  \S~5]{Folland89}} describes
all unitary irreducible representations of \(\Space{H}{n}\)
parametrised up to equivalence by two classes of
orbits~\eqref{eq:co-adjoint-orbits-inf}%
: 
  \begin{itemize}
  \item The infinite dimensional representations by transformation
    \(\uir{\myhbar}\)~\eqref{eq:stone-inf} for \(\myhbar \neq 0\) in
    Fock~\textup{\cite{Folland89,Howe80b}} space
    \(\FSpace{F}{2}(\orbit{\myhbar})\subset\FSpace{L}{2}(\orbit{\myhbar})\)
    of null solutions of Cauchy--Riemann type operators~\cite{Kisil02e}.
  \item The one-dimensional representations 
    which drops out from~\eqref{eq:stone-inf} for \(\myhbar =0\):
    \begin{equation}
      \label{eq:stone-one}
      \uir{(q,p)}(s,x,y): c \mapsto e^{-2\pi \rmi(qx+py)}c.
    \end{equation}
  \end{itemize}


Commutative representations~\eqref{eq:stone-one} are oftenly
forgotten, however their union naturally (see the appearance of
Poisson bracket in~\eqref{eq:Poisson}) act as the classic
\emph{phase space}:
\(
  \orbit{0}=\bigcup_{(q,p)\in\Space{R}{2n}} \orbit{(q,p)}.
\)

\subsection{Convolutions (observables) on $\Space{H}{n}$ and commutator}
\label{sec:conv-algebra-hg}

Using  a left invariant measure \(dg\) on \(\Space{H}{n}\) the linear space
\(\FSpace{L}{1}(\Space{H}{n},dg)\)  can be upgraded 
to an algebra with the convolution:
\begin{eqnarray}
  (k_1 * k_2) (g) &=& \int_{\Space{H}{n}} k_1(g_1)\,
  k_2(g_1^{-1}g)\,dg_1  .
  \label{eq:de-convolution}
\end{eqnarray}
Convolutions on \(\Space{H}{n}\) are \emph{observables} in
\(p\)-mechanic~\cite{Kisil96a,Kisil02e}.
Inner \emph{derivations} \(D_k\) of the convolution algebra
\(\FSpace{L}{1}(\Space{H}{n})\) are given 
by the \emph{commutator}:
\begin{eqnarray}
  D_k: f \mapsto [k,f]&=&k*f-f*k   \label{eq:commutator}
  \int_{\Space{H}{n}} k(g_1)\left( f(g_1^{-1}g)-f(gg_1^{-1})\right)\,dg_1.
\end{eqnarray}
A unitary representation \(\uir{\myhbar} \) of \(\Space{H}{n}\) extends
 to \(\FSpace{L}{1}(\Space{H}{n} ,dg)\):
\begin{equation}
  \fl
  \uir{\myhbar} (k) = \int_{\Space{H}{n}} k(g)\uir{\myhbar}  (g)\,dg .
  \label{eq:rho-extended-to-L1}
\end{equation}
Thus \(\uir{\myhbar} (k)\) for a fixed \(\myhbar \neq 0\) depends only
on \(\hat{k}_s(\myhbar,x,y)=\int k(s,x,y)\,e^{-2\pi\rmi hs}\,ds\)---the
partial Fourier transform \(s\mapsto \myhbar\) of \(k(s,x,y)\).
Consequently the representation of commutator~\eqref{eq:commutator}
depends only on:
\begin{eqnarray}
  \fl
    [k',k]\hat{_s}
 &=&   2 \rmi \! \int_{\Space{R}{2n}} \sin({\pi\myhbar}
 (xy'-yx'))\,\label{eq:repres-commutator}
 \hat{k}'_s(\myhbar ,x',y')\,
 \hat{k}_s(\myhbar ,x-x',y-y')\,dx'dy', 
\end{eqnarray}
which is exactly the Moyal bracket~\cite{Zachos02a} for the full
Fourier transforms of \(k'\) and \(k\). Also it vanishes for
\(\myhbar=0\) as can be expected from the commutativity of
representations~\eqref{eq:stone-one}.

\subsection{$p$-Mechanical bracket on $\Space{H}{n}$}
\label{sec:p-mechanical-bracket}

An antiderivative \(\anti\) is a scalar multiple of a right inverse
operator to the vector field
\(S\in\algebra{h}^n\)~\eqref{eq:h-lie-algebra}: 
\begin{equation}
  S\anti=4\pi^2 I, \textrm{ or }
  \label{eq:def-anti}
  \anti e^{ 2\pi\rmi \myhbar s}=\left\{ 
    \begin{array}{ll}
      \frac{2\pi}{\rmi\myhbar} \strut e^{2\pi\rmi\myhbar s}, & \textrm{if } \myhbar\neq 0,\\
      4\pi^2 s, & \textrm{if } \myhbar=0.
    \end{array}
    \right. 
\end{equation} 
It can be extended by the linearity to
\(\FSpace{L}{1}(\Space{H}{n})\). We introduce \(p\)-mechanical
bracket~\cite{Kisil00a,Kisil02e} on \(\FSpace{L}{1}(\Space{H}{n})\) as
a modified commutator of observables:
\begin{equation}
    \label{eq:star-and-brackets}
    \ub{k_1}{k_2} = (k_1*k_2-k_2*k_1)\anti.
\end{equation}
Then from~\eqref{eq:rho-extended-to-L1} one gets
\(\uir{\myhbar}(\anti k)=(i\myhbar)^{-1}\uir{\myhbar}(k)\) for
\(\myhbar\neq 0\). Consequently the modification
of~\eqref{eq:repres-commutator} for \(\myhbar\neq0\) is only slightly
different from the original one:
\begin{eqnarray}
  \fl
    \ub{k'}{k}\!\hat{_s}
    &=&   \int_{\Space{R}{2n}}
    \frac{2\pi}{\myhbar}\sin(\pi\myhbar  (xy'-yx'))\,\label{eq:repres-ubracket}
    \hat{k}'_s(\myhbar ,x',y')\,
    \hat{k}_s(\myhbar ,x-x',y-y') \,dx'dy'. 
\end{eqnarray}
However the last expression for \(\myhbar=0\) is significantly
distinct from~\eqref{eq:repres-commutator}, which vanishes as noted
above. From the natural assignment
\(\frac{4\pi}{\myhbar}\sin(\pi\myhbar (xy'-yx'))=4\pi^2(xy'-yx')\) for
\(\myhbar=0\) we get the Poisson bracket for the Fourier transforms
of \(k'\) and \(k\) defined on \(\orbit{0}\)
:
\begin{equation}
  \label{eq:Poisson}
    \uir{(q,p)}\ub{k'}{k} = \frac{\partial \hat{k}'}{\partial q}
    \frac{\partial \hat{k}}{\partial p}
    -\frac{\partial \hat{k}'}{\partial p} \frac{\partial \hat{k}}{\partial q}.
\end{equation}
Furthermore the dynamical equation~\cite{Kisil00a,Kisil02e} 
\begin{equation}
  \label{eq:p-dynamics}
  \dot{f}=\ub{H}{f} 
\end{equation}
based on the bracket~\eqref{eq:star-and-brackets} with a Hamiltonian
\(H(g)\) for an observable \(f(g)\) is
reduced~\cite{Kisil00a,Kisil02e} to Moyal's and Poisson's equations by
\(\uir{\myhbar}\) with \(\myhbar\neq 0\) and \(\myhbar= 0\)
correspondingly.  The same connections are true for the solutions of
these three equations, see~\cite{Kisil00a,Kisil02e,BrodlieKisil03a}.

\section{Mixed Quantum-Classical Bracket}
\label{sec:mixed-quant-class}

\subsection{A nilpotent group with two dimensional centre}
\label{sec:nilpotent-lie-group}
To derive quantum-classical bracket we again use the
``quantum-classical'' group \(\Space{D}{n}=\Space{H}{n} \oplus 
\Space{H}{n}\)~\cite{Prezhdo-Kisil97}.  This is a step 2
nilpotent Lie group of the (real) dimension \(4n+2\).  The group law is
given by the formula:
\begin{eqnarray}
\quad(g_1;g_2)*(g'_1;g'_2) = (g_1*g'_1; g_2*g'_2), 
\end{eqnarray} 
where \(g^{(\prime)}_i = (s^{(\prime)}_i, x^{(\prime)}_i,
y^{(\prime)}_i)\in \Space{H}{n}\), \(i=1,2\) and products \(g_i*g'_i\)
are the same as in
~\eqref{eq:H-n-group-law}. 

The group \(\Space{D}{n}\) has a two-dimensional centre
\(\Space{Z}{}=\{(s_1,0,0;s_2,0,0) \such s_1,s_2\in\Space{R}{}\}\).
The irreducible representations of a nilpotent group \(\Space{D}{n}\)
are induced~\cite[\S~13.4]{Kirillov76}, \cite[\S~6.2]{MTaylor86} by
the characters of the centre:
\(
  \mu: (s_1,s_2)\mapsto \exp(-2\pi\rmi(\myhbar_1 s_1 + \myhbar_2 s_2)). 
\) 
For \(\myhbar_1\myhbar_2\neq 0\) the
induced representation coincides with the irreducible representation
of \(\Space{H}{n+n}\):
\begin{equation}
  \uir{(\myhbar_1;\myhbar_2)}(g_1;g_2)
  =    \uir{\myhbar_1}(g_1)\,  \uir{\myhbar_2}(g_2) \label{eq:rho_hh}
\end{equation} 
This corresponds to \emph{purely quantum} behavior of both sets of
variables \((x_1, y_1)\) and \((x_2, y_2)\). The trivial character
\(\myhbar_1=\myhbar_2=0\) gives the family of one-dimensional
(\emph{purely classical}) representations parametrised by
points of \(\Space{R}{4n}\):
\begin{equation}\label{eq:rho_00}
  \uir{(q_1,p_1;q_2,p_2)}(s_1,x_1,y_1;s_2,x_2,y_2)
  = e^{ -2\pi\rmi (x_1 p_1 + y_1 q_1 + x_2 p_2 + y_2 q_2)}
\end{equation}

These cases for \(\Space{H}{n}\) were described above and studied in
details in~\cite{Kisil00a,Kisil02e,BrodlieKisil03a}. A new situation
appears when \(\myhbar_1\neq 0\) and \(\myhbar_2=0\) corresponding to
quantum behavior for \((x_1, y_1)\) and classical behavior for \((x_2,
y_2)\). The choice \(\myhbar_1= 0\), \(\myhbar_2 \neq 0\) swaps the
quantum and classical parts.
The quantum-classical representation 
is given by 
\begin{eqnarray}
  \uir{(\myhbar;q,p)}(g_1; g_2)
  &=& \uir{\myhbar}(g_1)\uir{(q,p)}(g_2) 
  = \uir{\myhbar}(s_1,x_1,y_1) e^{-2\pi\rmi(qx_2+py_2)}, \label{eq:rho_h0}
\end{eqnarray}
where \(q,p\in \Space{R}{n}\) and \(\myhbar\in \Space{R}{}\setminus
\{0\} \).  In this representation a convolution (observable) on
\(\Space{D}{n}\) generates a function on the classic phase space
\(\Space{R}{2n}\) with values in space of quantum operators acting of
\(L_2(\Space{R}{n})\), cf.~\cite{Aleksandrov81}, or explicitly:
\begin{eqnarray}{}
\fl \uir{(\myhbar;q,p)}k &=&  \int_{\Space{D}{n}}
k(g_1;g_2)\,\uir{\myhbar}(g_1) e^{-2\pi\rmi(qx_2+py_2)}\,dg_1\,dg_2
  \label{eq:rho_h0_conv}\\
&=& 
  \int_{\Space{H}{n}} \hat{k}_2(s_1, x_1, y_1;0,q,p)\,\uir{\myhbar}(s_1,x_1,y_1)
\,dg_1
\nonumber 
\end{eqnarray}
where \(\hat{k}_2\) is the partial Fourier transform of \(k\) with
respect to variables \((s_2, x_2, y_2)\mapsto (h,q,p)\).

\subsection{The Mixed Bracket}
\label{sec:mixed-bracket}
We define \(p\)-bracket in the case of \(\Space{D}{n}\) similarly
to~\eqref{eq:star-and-brackets}. Although this is not a unique option,
some other similar definitions may be of interest as well.
\begin{defn}
  \label{de:p-brackets}
  The \emph{\(p\)-\-mechanical bracket} of two convolutions
  (observables) \(k_1(g_1;g_2)\) and
  \(k_2(g_1;g_2)\) on the group \(\Space{D}{n}\) is defined as
  follows: 
  \begin{equation}\label{eq:p-brackets}
    \ub{k_1}{k_2}=(k_1*k_2-k_2*k_1) (\anti_1+\anti_2),
  \end{equation} where \(*\) denotes the group convolution on
  \(\Space{D}{n}\). \(\anti_1\) and \(\anti_2\) are
  antiderivatives with respect to the  variable \(s_1\) and \(s_2\)
  correspondingly, cf.~\eqref{eq:def-anti}. 
\end{defn} 
Consistence of this definition, cf.~\cite{CaroSalcedo99},
is given by:
\begin{lem}
  \label{le:consistency}
  The \(p\)-\-mechanical bracket~\eqref{eq:p-brackets} is linear,
  antisymmetric, satisfy Leibniz and Jacoby
  identities. Moreover \(p\)-mechanical bracket with a Hamiltonian has
  the dimensionality of time derivative. 
\end{lem}
We define \emph{\(p\)-mechanisation}~\cite{Kisil02e} of a classical
observable \(f(q, p)\) is given by the Weyl (symmetrized)
calculus~\cite{Folland89} defined on the generators as follows:
\begin{equation}
  \label{eq:mechanisation}
  q_j \mapsto Q_j=\delta'_{x_j}(g_1;g_2), \qquad p_1 \mapsto
   P_j=\chi'_{s_k}(s_1+s_2)*\delta'_{y_j}(g_1;g_2), \quad
   j=1,2 \textrm{ and } k=3-j,
\end{equation}
where \(\delta'_{z}\) is the derivative of the Dirac delta function
with respect to the variable \(z\) and \(\chi'_{s_k}\) is the
derivative of the Heaviside
step function such that \(\chi'_z(z)=\delta(z)\). Using the identity
\begin{eqnarray}
  \label{eq:canon-commut}
  \int_{\Space{R}{}} \chi'_{s_k}(s_1+s_2)\,e^{-2\pi\rmi(\myhbar_1 s_1
    + \myhbar_2 s_2)}\,dz=\frac{\myhbar_k}{\myhbar_1+\myhbar_2}
  \quad\textrm{ we get that }\quad \ub{Q_i}{P_j}=\delta_{ij}I,
\end{eqnarray}
and all other brackets vanish.
Representations of distributions~\eqref{eq:mechanisation} and the
bracket~\eqref{eq:p-brackets} are: 
\begin{equation}
  \label{eq:representations}
  \begin{array}{c||c|c|r|c}
     &  \uir{(\myhbar_1;\myhbar_2)}  &  \uir{(\myhbar;q, p)}
     &  \intersect{\partial_{\myhbar_2}\uir{(\myhbar;q,  p)}}{\myhbar_2=0}\qquad
     &  \uir{(q_1,p_1;q_2,p_2)} \\
    \hline\hline
     Q_j\strut & \frac{\strut}{\strut}\partial_{x_j}-\frac{\rmi\myhbar_j}{2}y_j & 
    \begin{array}{c}
      \partial_{x_1}-\rmi\myhbar y_1/2 \\ 
      \rmi q\\
    \end{array} &  
    \begin{array}{ll}
      0,&\textrm{if } j=1\\
      \partial_p/2, &\textrm{if } j=2
    \end{array} & q_j\\
    \hline
    P_j\strut &  \frac{\strut}{\strut}\frac{\myhbar_k}{\myhbar_1
      +\myhbar_2}\left(\partial_{y_j}+\frac{\rmi \myhbar_j}{2}
      x_j\right) &  
    \begin{array}{c}
      0\\
      \rmi p\\
    \end{array} &  
    \begin{array}{ll}
      \partial_{y_1}/\myhbar+\rmi x_1/2,&\textrm{if } j=1\\
      -\rmi p/\myhbar-\partial_q/2, &\textrm{if } j=2
    \end{array} &  
    p_j \\
    \hline
    \ub{K_1}{K_2}  &   \frac{\strut}{\strut}\left(\frac{1}{\rmi\myhbar_1} + \frac{1}{\rmi\myhbar_2} \right)
      [K_1,K_2]  & \multicolumn{2}{c}{ [K_1,K_2]_{qc}} \vline& 
      \displaystyle \{\hat{k}_1,\hat{k}_2\}
  \end{array}
\end{equation}
  where the bracket \([\cdot,\cdot]_{qc}\) in the
  quantum-classical case is defined by the expression:
  \begin{eqnarray}
    {}[K_1,K_2]_{qc}&=& \frac{1}{ \rmi\myhbar} [K_1,K_2] 
    +\frac{1}{2}\left(\{K_1,K_2\}-\{K_2,K_1\}\right)
    -\intersect{\rmi\partial_{\myhbar_2} 
      [K_1,K_2]}{\myhbar_2=0}. \label{eq:qc-brackets}
  \end{eqnarray}
  Calculations of the two first terms in~\eqref{eq:qc-brackets} is
  similar to \(p\)-bracket~\cite{Kisil00a,Kisil02e}, the third term
  is:
  \begin{eqnarray*}
    \lefteqn{\int\limits_{\Space{D}{n}}\int\limits_{\Space{D}{n}}\!
     (k_1(g'_1;g_2')\,k_2(g^{\prime-1}_1g_1;g_2'')
     -k_2(g'_1;g_2')\,k_1(g^{\prime-1}_1g_1;g_2''))}\qquad\\
   &&\times
   (s_2''+s_2')\,e^{-2\pi\rmi(qx_2'+py_2'+qx_2''+py_2'')}
   \,dg'_2\,dg''_2 dg'_1\, \uir{h}(g_1)\,dg_1.
  \end{eqnarray*}
  The complete derivation will be given elsewhere. The derivative
  \(\partial_{\myhbar_2}\) in~\eqref{eq:qc-brackets} highlights the
  important difference between Aleksandrov's~\cite{Aleksandrov81} and
  our approach: \emph{quantum-classical observables are not operator
    valued functions on the classical phase space but rather first
    jets~\textup{\cite{Olver93}} of such functions}. This explain the
  appearance of the fourth column in~\eqref{eq:representations}.

By algebraic inheritance~\textup{\cite{Kisil00a}} the quantum-classic
bracket~\eqref{eq:qc-brackets} enjoys all the properties from
Lem.~\ref{le:consistency}. Moreover quantum-classical bracket
coincides with the Moyal bracket for purely quantum observables and
the Poisson bracket for purely classical ones.
  Let a \(p\)-mechanical observable \(f(t;g_1;g_2)\), which is a function on
  \(\Space{R}{}\times\Space{D}{n}\), be a solution of the equation:
  \begin{equation}\label{eq:universal1}
    \frac{d}{dt} f(t;g_1;g_2)= \ub{f}{H}
  \end{equation}
  with a Hamiltonian \(H(g_1,g_2)\) on \(\Space{D}{n}\). Then
  \(f(t;g_1;g_2)\) provide consistent dynamics (in the sense
  of~\cite{CaroSalcedo99}) under either representation
  \eqref{eq:rho_hh}--\eqref{eq:rho_h0}. 
\begin{example}[Dynamics with two different Planck's constants, cf.~\textup{\cite{Sahoo04}}]
  Let \(p\)-mechanical Hamiltonian is defined by such a distribution
  on \(\Space{D}{n}\) (see definitions~\eqref{eq:mechanisation}):
  \begin{eqnarray}
    \label{eq:rot-hamiltonian}
   H&=&Q_1P_2-Q_2P_1
    = \chi'_{s_1}(s_1+s_2)*\delta^{(2)}_{x_1,y_2}(g_1;g_2)
    -\chi'_{s_2}(s_1+s_2)*\delta^{(2)}_{x_2,y_1}(g_1;g_2).
  \end{eqnarray}
  In the classic-classical representation~\eqref{eq:rho_00} it
  produces (see the last column of~\eqref{eq:representations})
  the quadratic Hamiltonian \(H_{cc}=q_1p_2-q_2p_1\), which generates a simple
  rotational dynamic:
  \begin{eqnarray}
    q_1(t)= \cos t\, q_1(0) + \sin t\, q_2(0), && 
    q_2(t)= -\sin t \,q_ 1(0)+ \cos t\, q_2(0), \label{eq:class-dyn-1}\\
    p_1(t)= \cos t\, p_1(0) + \sin t\, p_2(0), && 
    p_2(t)= -\sin t\,p_1(0) + \cos t\, p_2(0). \label{eq:class-dyn-2}
  \end{eqnarray}
  In the quantum-quantum representation~\eqref{eq:rho_hh} defined by
  two Planck's constants \(\myhbar_1\) and \(\myhbar_2\)
  (\(\myhbar_1\myhbar_2\neq 0\)) the Hamiltonian becomes (see the first
  column of~\eqref{eq:representations}):
  \begin{displaymath}
    \textstyle H_{qq} =   \frac{\myhbar_1}{\myhbar_1 +\myhbar_2}
    \left(\partial_{x_1}-\frac{\rmi\myhbar_1}{2}y_1\right)
    \left(\partial_{y_2}+\frac{\rmi \myhbar_2}{2} x_2\right)
    -  \frac{\myhbar_2}{\myhbar_1 +\myhbar_2} 
    \left(\partial_{x_2}-\frac{\rmi\myhbar_2}{2}y_2\right)
    \left(\partial_{y_1}+\frac{\rmi \myhbar_1}{2} x_1\right).
  \end{displaymath}

  The dynamic from the bracket \(\left(\frac{1}{\rmi\myhbar_1} +
    \frac{1}{\rmi\myhbar_2} \right) [H_{qq},f]\)
  in~\eqref{eq:representations} is the coordinate map on 
  \(\Space{D}{n}\):
  \begin{eqnarray}
    x_1(t)= \cos t\, x_1(0) + \sin t\, x_2(0),\qquad\  &&
    \quad 
    x_2(t)=  -\sin t \, x_1(0)+ \cos t\, x_2(0), \label{eq:quant-dyn-1}\\  
    \myhbar_2 y_1(t)=\myhbar_2 \cos t\, y_1(0)+\myhbar_1 \sin t\,
    y_2(0), && 
    \myhbar_1 y_2(t)=  -\myhbar_2\sin t\, y_1(0) + \myhbar_1\cos t\,
    y_2(0). \label{eq:quant-dyn-2} 
  \end{eqnarray}
  For \(\myhbar_1=\myhbar_2\) this coincides with the standard
  quantisation of the classical
  dynamics~\eqref{eq:class-dyn-1}--\eqref{eq:class-dyn-2}.
  
  The quantum-classic Hamiltonian is the \(1\)-jet (see the middle column
  of~\eqref{eq:representations}):
  \begin{displaymath} \textstyle
    H_{qc}=\left(\rmi (\partial_{x_1}-\frac{\rmi\myhbar}{2} y_1)p, 
      \quad
      -\frac{\rmi}{\myhbar}
      \left(\partial_{x_1}-\frac{\rmi\myhbar}{2}y_1\right)
        \left(p-\frac{\rmi \myhbar}{2} \partial_q\right)
       -\frac{\rmi}{\myhbar}q \left(\partial_{y_1}+\frac{\rmi
           \myhbar}{2} x_1\right)\right). 
  \end{displaymath}
  Note, that Aleksandrov's bracket~\eqref{eq:aleksandrov-brackets} of
  \(H_{qc}\) with \(\uir{(\myhbar;q,p)}(Q_1)\) vanish and thus do not
  generate any dynamics for this observable. However \([H_{qq},
  \uir{(\myhbar_1;\myhbar_2)}(Q_1)]=
  \frac{\rmi\myhbar_1\myhbar_2}{\myhbar_1 +
    \myhbar_2}\uir{(\myhbar_1;\myhbar_2)}(Q_2)\) and thus the third
  term in the bracket~\eqref{eq:qc-brackets} of \(H_{qc}\) and
  \(\uir{(\myhbar;q,p)}(Q_1)\) is equal to \(\uir{(\myhbar;q,p)}(Q_2)
  = \rmi q\) (this is also the value of the entire
  bracket~\eqref{eq:qc-brackets}).  Together with the value of
  \([H_{qc}, \uir{(\myhbar;q,p)}(Q_2)]_{qc} =
  -\uir{(\myhbar;q,p)}(Q_1)\) this defines quantum-classic dynamics of
  coordinates as the partial Fourier transform \(x_2\mapsto q\) of the
  quantum-quantum coordinate map~\eqref{eq:quant-dyn-1}.  

  Similarly we calculate that \([H_{qc},
  \uir{(\myhbar;q,p)}(P_1)]_{qc} = \uir{(\myhbar;q,p)}(P_2) = \rmi p\)
  and \([H_{qc}, \uir{(\myhbar;q,p)}(P_2)]_{qc} =
  -\uir{(\myhbar;q,p)}(P_1)\). Note that \(\uir{(\myhbar;q,p)}(P_1)\)
  is the \(1\)-jet with the value \((0,
  \frac{1}{\myhbar}\partial_{y_1}+\frac{\rmi}{2} x_1)\) according to
  the~\eqref{eq:representations} and the quantum-classic bracket
  depends from its both components. The quantum-classic dynamic of
  momenta is obtained from~\eqref{eq:quant-dyn-2} by
  prolongation~\cite{Olver93} into the \(1\)-jet space with respect to
  the variable  \(\myhbar_2\) at point \(\myhbar_2=0\).
\end{example}

\section{Conclusion}
\label{sec:conclusion}
The sum~\eqref{eq:aleksandrov-brackets} of first two terms in~\eqref{eq:qc-brackets} 
was proposed~\cite{Aleksandrov81,BoucherTaschen88} as a version of
quantum-classical bracket. It was also obtained by approximation
arguments within \(p\)-\-mechanical approach
in~\cite{Prezhdo-Kisil97} as a part of the true bracket unknown
at that time. However the expression~\eqref{eq:aleksandrov-brackets}
violates the Jacobi identity and Leibniz rule (i.e. is not a
derivative), as a consequence it could not be used for a consistent
dynamic equation~\cite{CaroSalcedo99}. Our new
bracket~\eqref{eq:qc-brackets} has one extra term which makes it
satisfactory to this end. This term is of an analytical nature (i.e.
involves a derivative in Planck's constant) and is hard to guess
from algebraic manipulations with the quantum commutator and
Poisson's bracket. For the same reasons our
bracket~\eqref{eq:qc-brackets} are immunised against the ``no-go''
theorem of the type proved
in~\cite{Salcedo96,CaroSalcedo99,Sahoo04}.
We present an example of a
dynamics~\eqref{eq:quant-dyn-1}--\eqref{eq:quant-dyn-2}, which mixes
two quantum sectors with different Planck's constants, and demonstrate
the quantum-classic dynamics in the \(1\)-jet space. 

I am grateful to Prof.~G.~Ingold and anonymous 
referees who helped to improve this letter.\vspace{-4mm}

\newcommand{\noopsort}[1]{} \newcommand{\printfirst}[2]{#1}
  \newcommand{\singleletter}[1]{#1} \newcommand{\switchargs}[2]{#2#1}
  \newcommand{\irm}{\textup{I}} \newcommand{\iirm}{\textup{II}}
  \newcommand{\vrm}{\textup{V}}
  \providecommand{\cprime}{'}\providecommand{\arXiv}[1]{\eprint{http://arXiv.o%
rg/abs/#1}{arXiv:#1}}
\providecommand{\bysame}{\leavevmode\hbox to3em{\hrulefill}\thinspace}
\providecommand{\MR}{\relax\ifhmode\unskip\space\fi MR }
\providecommand{\MRhref}[2]{%
  \href{http://www.ams.org/mathscinet-getitem?mr=#1}{#2}
}
\providecommand{\href}[2]{#2}

\end{document}